\documentclass[%
 aip,
rsi,%
 amsmath,amssymb,
reprint,%
prl,
]{revtex4-1}

\usepackage{graphicx}
\usepackage{dcolumn}
\usepackage{bm}

\begin{document}

\preprint{AIP/123-QED}

\title[Scaling Law in Laser Cooling on Narrow-Line Optical Transitions] {Scaling Law in Laser Cooling on Narrow-Line Optical Transitions}

\author{O.N.Prudnikov}
\email{oleg.nsu@gmail.com}
\author{R.Ya Il'enkov}%
\author{A.V. Taichenachev}%
\affiliation{
Institute of Laser Physics, 630090, Novosibirsk, Russia 
}%
\affiliation{
Novosibirsk State University, 630090, Novosibirsk, Russia 
}%

\author{V.I. Yudin}%
\affiliation{
Institute of Laser Physics, 630090, Novosibirsk, Russia 
}%
\affiliation{
Novosibirsk State University, 630090, Novosibirsk, Russia 
}%
\affiliation{Novosibirsk State Technical University, 630073,
Novosibirsk, Russia }

\date{\today}

\begin{abstract}
In this paper laser cooling of atoms with a narrow-line optical
transition, i.e. in regimes of quantum nature of laser-light
interactions resulting in a significant recoil effect, is studied.
It is demonstrated that a minimum laser cooling temperature for
two-level atom in standing wave reached for red detuning close to 3
recoil frequency greatly different from the theory used for a
semiclassical description of Doppler cooling. A set of dimensionless
parameters uniquely characterizing the time evolution and the steady
state of different atoms with narrow-line optical transitions in the
laser field is introduced. The results can be used for analysis of
optimal conditions for laser cooling of atoms with narrow lines such
as $Ca$, $Sr$, and $Mg$, which are of great interest for atomic
clocks.
\end{abstract}



\maketitle

Nowadays deep laser cooling of neutral atoms is routinely used for
wide range of modern quantum physics investigations, including
metrology, atom optics, and quantum degeneracy studies. There exist
well-known techniques for laser cooling below the Doppler limit,
such as sub-Doppler polarization gradient cooling
\cite{Dalibard1989,Castin1991,Pru1999,Pru2004}, velocity selective
coherent population trapping
\cite{Aspect88,Aspect1989,Prudnikov2003}, or Raman cooling
\cite{Kasevich1992,ctannoudji1995} restricted to atoms with energy
levels degenerated over angular momentum  or hyperfine structure.
However, these techniques can not be applied to atoms with a single
ground state, such as $^{24}$Mg, $^{40}$Ca, $^{88}$Sr, $^{174}$Yb,
which are of great interest for atomic clocks.

For atoms with a single nondegenerated ground state the well-known
theory based on a semiclassical approach predicts so-called Doppler
 laser cooling temperature $k_B T_D
\approx \hbar \gamma/2$, with the natural linewidth $\gamma$ of
optical transition \cite{kaz_book,metcalf}. One way of reaching a
deeper cooling for these atoms is to use a narrow-line optical
transition (clock transition) with a smaller natural linewidth
$\gamma$. However, the basic semiclassical theory is not valid,
since the main requirement, called the semiclassical limit
$\omega_R/\gamma \ll 1$ (where $\omega_R = \hbar k^2/2M$ is the
recoil frequency describing the energy acquired by an atom at rest
due to spontaneous emission or absorption of a light field photon
with momentum $\hbar k$) is violated, that was clearly shown in
\cite{castin1989} for atom in $\sigma_+-\sigma_-$ light field
configuration.

 There are several experimental
realizations of laser cooling of $^{174}Yb$ atoms (at the optical
transition $^1S_0 \to ^3P_1$, $\lambda = 556$ nm  with
$\omega_R/\gamma = 0.02$)\cite{Maruyama2003} and $^{88}Sr$ atoms (at
the transition $^1S_0 \to ^3P_1$, $\lambda = 689$ nm with
$\omega_R/\gamma = 0.635$)\cite{Katori1999,Strelkin2015}. For the
cooling of $^{88}Sr$ atoms the authors uses broadband light sources
to increase velocity range of atoms much large than the typical
Doppler velocity range $v_D = \gamma/k$ \cite{Ertmer1989}.
 Direct cooling with the use
of narrow lines of $^{40}Ca$ (for $\lambda = 657$ nm with
$\omega_R/\gamma = 32.3$) and $^{24}Mg$ (for $\lambda = 457$ nm with
$\omega_R/\gamma = 1100$) does not seem possible due to a much
larger recoil energy then the transition linewidth. Estimated
Doppler limit temperatures for laser cooling at the narrow-line
optical transition $^1S_0 \to ^3P_1$ for various atoms and values of
the semiclassical parameter $\omega_R/\gamma$ are presented in Table
\ref{tab1}.
\begin{table}
\caption{Semiclassical estimation of temperatures of laser cooling
on intercombination lines $^1S_0 \to ^3P_1$.}\label{tab1}
\setlength\extrarowheight{2.5pt}
    \begin{tabular}{|p{50pt}|p{50pt}|p{50pt}|p{50pt}|}
    \hline
    Atom & $T_D$ & $\lambda $& $\omega_R/\gamma$ \\ \hline
    $^{200}Hg$ & 32 $\mu K$ & 254 nm & 0.01 \\
    $^{174}Yb$ & 4.5 $\mu K$ & 556 nm & 0.02 \\
    $^{88}Sr$ & 0.17 $\mu K$ & 689 nm& 0.635 \\
    $^{40}Ca$ & 10 $nK $& 657 nm & 32.3 \\
    $^{24}Mg$ & 0.75 $nK $& 457 nm & 1100  \\ \hline
\end{tabular}
\end{table}

Note that deep laser cooling of $Ca$ atoms at the narrow-line
optical transition was made using ``quenching cooling'' techniques
\cite{holberg2001,Rasel2001,Rasel2003}. These techniques involve the
use of an additional light field, which allows increasing the
effective optical transition linewidth $\gamma_{eff} \gg \gamma$ to
``return back'' to the semiclassical conditions
$\omega_R/\gamma_{eff} \ll 1$ for the standard laser cooling regime.
However, to our knowledge, no significant progress for $^{24}$Mg
atoms has been achieved so far.

In this letter, we present a quantum theory of laser cooling in
standing wave for regimes far beyond the semiclasical limit, i.e.
for $\omega_R/\gamma \geq 1$. This allows to clarify the laser
cooling mechanisms with narrow-line optical transitions and estimate
the optimal parameters for minimum cooling temperatures and cooling
times especially for enough strong light field intensity resulting
to ``power broadening'' regime of laser cooling. We find the laser
cooling dynamics and steady-state momentum distribution can be
uniquelly characterized by set of dimentionless parameters for
various atoms with different $\omega_R/\gamma$ ratio that results to
a ``scaling law'' in laser cooling.

Let us consider the typical scheme of laser cooling of atoms in a
light field formed by two counterpropagating monochromatic waves
(see Fig.1) resulting to standing wave with intensity modulation.
The light field is close to the resonance of an atomic optical
transition with the natural linewidth $\gamma$. The behavior of
atoms in the laser light field can be described by density matrix
master equation with taking into account the quantum recoil effects
in the processes of absorption and emission of photons

\begin{figure}[h]
\begin{center}
\includegraphics[width= 2.8 in]{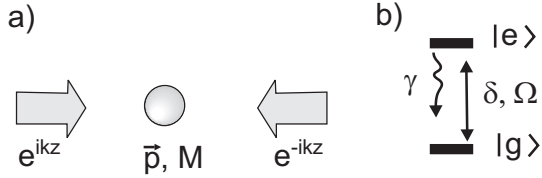}
\end{center}
\caption{\em Scheme of laser cooling in a standing light wave (a).
Scheme of interaction of resonant laser light with atom energy
levels (b).}
\end{figure}

\begin{equation}\label{master}
\frac{d}{dt}{\hat \rho} = -\frac{i}{\hbar}\left[ {\hat H}, {\hat
\rho}\, \right] + {\hat \Gamma}\left\{{\hat \rho}\right\} \, ,
\end{equation}
where ${\hat H}$ is the Hamiltonian and ${\hat \Gamma}\left\{{\hat
\rho}\right\}$ is the relaxation operator due to spontaneous
emission (see, for example, \cite{kaz_book}). The density matrix
contains all information about the internal and external states of
atoms and the correlations in the processes of atom-light
interactions.

The Hamiltonian in a resonant light field in the rotating wave
approximation (RWA) has the following form:
\begin{equation}
{\hat H} = \frac{{\hat p}^2}{2M} - \hbar \delta {\hat P}_e + {\hat
V}_{ed} \, .
\end{equation}
The first term describes the kinetic energy of atoms, ${\hat P}_e =
|e \rangle \langle e |$ is the projection operator to the excited
state $|e \rangle $, ${\hat V}_{ed} = \hbar \Omega/2 |e \rangle
\langle g | +h.c.$ is the atom-light coupling of the ground (g) and
excited (e) states in the electric dipole approximation, and
$\Omega$ is the Rabi frequency. In standing light field the Rabi
frequency has a spatial modulation $\Omega(z) = 4 \Omega_0
\cos^2(kz)$, where $\Omega_0$ is the Rabi frequency per one wave.
Here  $\delta = \omega-\omega_0$ is detuning of the laser light
frequency $\omega$  from the atom optical transition frequency
$\omega_0$.

First of all, let us emphasize the light field and atomic parameters
that define the time evolution and the steady state of atom kinetics
in the light field. The master equation (\ref{master}) is determined
by the following parameters of frequency dimension: $\gamma$,
$\delta$, $\Omega$, and  $\omega_R$.

The semiclassical regime is characterized by small recoil frequency
\begin{equation}\label{semiclassic}
\omega_R/\gamma \ll 1 \,,
\end{equation}
that is valid for major atoms with strong dipole optical transitions
where laser cooling was realized (see for example \cite{Adams1997}).

\begin{figure}[h]
\centerline{\includegraphics[width=3.5 in]{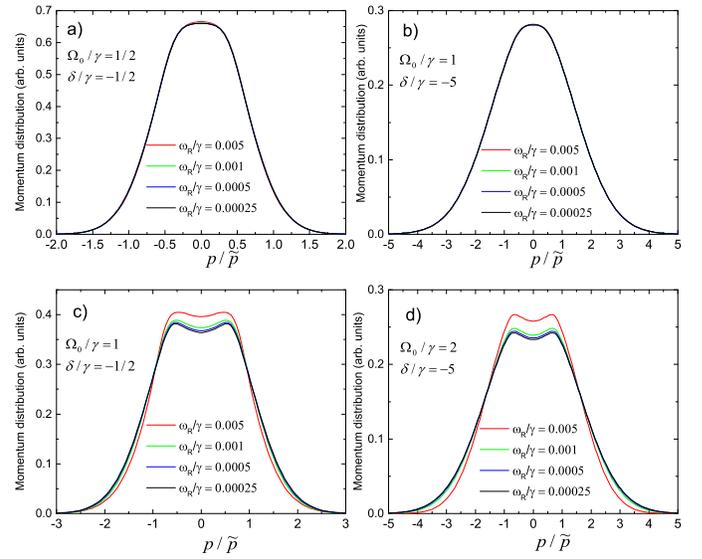}} \caption{\em
The steady-state momentum distribution function of cold atoms in the
semiclassical limit (\ref{semiclassic}). Field parameters:
$\Omega_0/\gamma = 1/2$, $\delta/\gamma = - 1/2$ (a),
$\Omega_0/\gamma = 1$, $\delta/\gamma = -5$ (b), $\Omega_0/\gamma =
1$, $\delta/\gamma = - 1/2$ (c), and $\Omega_0/\gamma = 2$,
$\delta/\gamma = -5$ (d).} \label{mom_mem_fig}
\end{figure}

Let us mention, within the framework of a two-level model (an atom
with nondegenerate ground state) the atoms steady state distribution
of laser cooling in the limit (\ref{semiclassic}) can be equally
described by only two dimentionless parameters:
\begin{equation}\label{semparameters}
\delta/\gamma, \,\,\, \Omega/\gamma.
\end{equation}
The cold atoms steady-state momentum distribution function for low
saturation $S=|\Omega|^2/(\delta^2+\gamma^2/4) \leq 1$ and red
detuning scales in these dimentionless parameters as
\begin{equation}
\label{momentum_st} {\cal F}_p = {\cal F}\left(\frac{p}{\tilde{p}},
\frac{\Omega}{\gamma}, \frac{\delta}{\gamma}\right), \,\,\,\,\,
\tilde{p}=\hbar k \sqrt{\gamma/\omega_R}
\end{equation}
that represents the scaling law in semiclassical limit
(\ref{semiclassic}). The momentum distribution for all atoms in the
semiclassical limit (\ref{semiclassic}) get equivalent from for
momentum expressed in Doppler momentum units $\tilde{p}$
(\ref{momentum_st}). In particular, the steady-state average kinetic
energy in the semiclassical limit scales in $\gamma$ units and is a
function of light field dimentionless parameters
(\ref{semparameters}):
\begin{equation}\label{En_sem}
E_{kin} =\langle p^2/2M \rangle = \hbar \gamma {\cal
F}_{S}(\delta/\gamma, \Omega/\gamma)
\end{equation}
The function ${\cal F}_{S}(\delta/\gamma,\, \Omega/\gamma)$ (see for
example \cite{kaz_book}) for two-level atom has the minimum $\simeq
1/6$ at $\delta/\gamma = -1/2$ for $\Omega/\gamma \rightarrow 0$, so
called Doppler cooling limit.

The momentum distribution on Fig.\ref{mom_mem_fig} is obtained by
numerical solution of density matrix equation (\ref{master}) with
the use of the method suggested by us in \cite{Pru2007,Pru2011}. The
momentum distribution function is well scaled for introduced set of
parameters, and in the limit of small ratio $\omega_R/\gamma$ the
difference between the curves becomes less noticeable in low
intensity (see Fig.\ref{mom_mem_fig}(a) and (b)). Here for the
Fig.\ref{mom_mem_fig}(c) and (d) we choose larger field intensity to
get difference between the curves more visible.

The evolution of the momentum distribution of an atomic cloud and
the laser cooling rate are determined by the slowest processes of
atom momentum distribution function modification due to interaction
with light field photons through the exchange of photon momentum.
The cooling rate of slow atoms with $p<\hbar k \gamma/\omega_R$ in
semiclassical limit (\ref{semiclassic}), can also be represented in
equivalent form for various atoms by introducing a dimensionless
time $t/t_S$
\begin{equation}\label{semtime}
t_S = \omega_{R}^{-1} \cdot \tau_S\left(\Omega/\gamma,\,
\delta/\gamma \right)\, ,
\end{equation}
that scales in $\omega_{R}^{-1}$ units and is a function of
dimensionles parameters $\Omega/\gamma$ and $\delta/\gamma$ only.

\begin{figure}[h]
\centerline{\includegraphics[width=2.9 in]{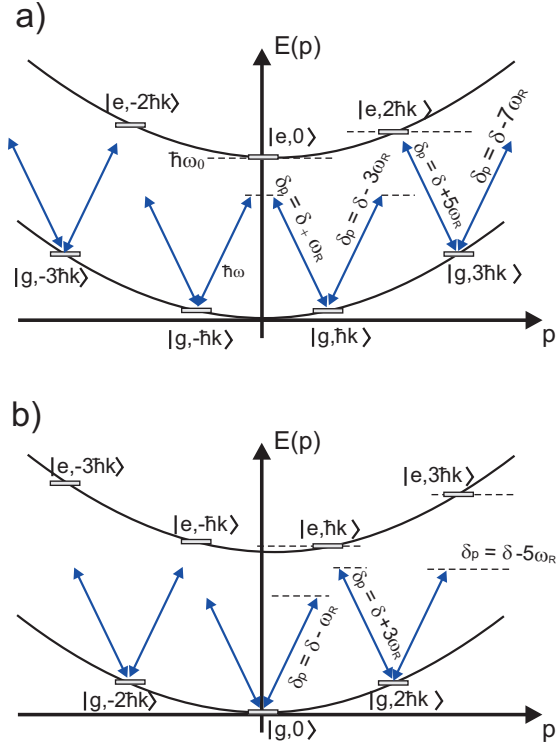}}
\caption{\em Schemes of atom light interaction for atoms with narrow
lines in momentum representation: for quasi-momentum $p_0 = \hbar k$
(a) and $p_0 = 0$ (b).} \label{P_levels}
\end{figure}

On a base of laser cooling theory in semiclassical limit from the
general relations (\ref{semparameters}) - (\ref{semtime}) one may
conclude that the narrow-line optical transitions with small natural
linewidth might be perspective for deep and fast laser cooling, been
considering same optimal field parameters as semiclassical theory
predicts. This naive picture is not correct. For the laser cooling
regimes far beyond the semiclassical limit (\ref{semiclassic}), the
recoil effects becomes essential as it was demonstrated in
\cite{dal1989} for $\sigma_+-\sigma_-$ field configuration. Here we
may also expect significant modification of the scaling law.

In the quantum regime being considered
\begin{equation}\label{quantum}
\omega_R \geq \gamma
\end{equation}
the simple scheme of two-level atom interaction with light standing
wave (Fig.1(b)) is significantly modified. The internal states can
be written in a momentum representation as a set of families with a
different momentum $p$ with a difference $\Delta p = 2 \hbar k$ in
the ground and excited states:
\begin{equation} \label{w_functionP}
|\psi(t) \rangle = \sum_n \alpha_n(t) |g,p_0+2 n \hbar k \rangle +
\beta_n(t) |e,p_0+ (2 n+1) \hbar k  \rangle \, .
\end{equation}
with quasi-momentum $p_0$ in the range $-\hbar k \leq p_0 \leq\hbar
k$. Coupling schemes for families for $p_0 = \hbar k$ and $p_0 = 0$
are shown in Fig. \ref{P_levels}. In the regime (\ref{quantum}) the
processes of induced absorption and emission of light field photons
are beyond the resonance contour $\gamma$, and have different
detuning:
\begin{equation}\label{deltaP}
\delta_p = \delta -\omega_R \left(1\mp 2p/\hbar k \right) \, ,
\end{equation}
depending on the atom momentum on the ground state, $p$. Here the
sign ``-'' is related to induced emission of the photon to
copropagating wave or absorption from counterpropagating wave, and
the sign ``+'' related to reverse processes.

Let us define the scaling law in the regime (\ref{quantum}). Here
the equivalent parameters universally describing laser cooling of
various atoms (\ref{semparameters}) and (\ref{semtime}), are no
longer valid. In this regime $\gamma$ is the smallest parameter and
the steady-state solution of the master equation is governed by
different set of  dimensionless parameters:
\begin{equation}\label{Q_parameters}
\delta/\omega_R, \,\, \Omega/\omega_R \, ,
\end{equation}
defining the equivalence of laser cooling of various atoms with the
use of the narrow-line transitions, that is, in the case of an
essential quantum recoil effect in the atom-light interaction. The
momentum distribution of atoms is a function of parameters scales
not in $\gamma$, but in $\omega_R$ units, contrary to semiclassical
limit:
\begin{equation}\label{mom_Q}
{\cal F}_p = {\cal F}\left(\frac{p}{\hbar k},
\frac{\Omega}{\omega_R}, \frac{\delta}{\omega_R}\right)
\end{equation}
and the average kinetic energy is determined by
\begin{equation}\label{en_q}
E_{kin} = \langle p^2/2M \rangle =\hbar \omega_R {\cal
F}_{Q}(\delta/\omega_R, \Omega/\omega_R)
\end{equation}
with ${\cal F}_{Q}(\delta/\omega_R, \Omega/\omega_R)$ is
dimensionless function of introduced parameters.

To confirm this, we perform a numerical simulation of the master
equation (\ref{master}) for various atoms ($Sr$, $Mg$, $Ca$) with
narrow-line optical transitions for a set of different parameters
(\ref{Q_parameters}).

The steady-state momentum distribution of atoms was obtained by
solving the master equation (\ref{master}) numerically by a method
developed by us in \cite{Pru2007,Pru2011}. The method proposed
allows one to obtain a direct solution for a density matrix
containing all information on internal and external states without
any restrictions and limitations.

\begin{figure}[t]
\centerline{\includegraphics[width=2.6 in]{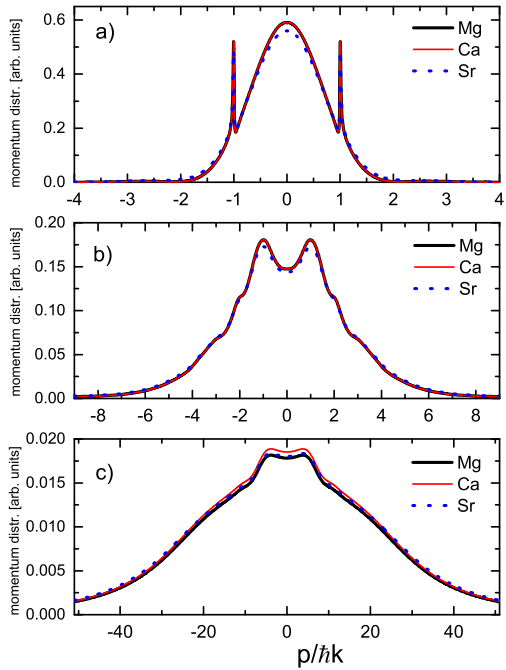}}
\caption{\em The steady-state momentum distribution of laser cooled
atoms for different set of equivalent parameters
(\ref{Q_parameters}). The light field detuning $\delta = -3
\omega_R$ and Rabi per one wave $\Omega_0 = 0.64 \,\omega_R$ (a),
$\Omega_0 = 6.4\,\omega_R$ (b), and $\Omega_0 = 64\,\omega_R$ (c).}
\label{fig_dist}
\end{figure}

Fig.\ref{fig_dist} shows the results of steady state momentum
distribution of various atoms ($^{88}Sr$, $^{24}Mg$, $^{40}Ca$) for
the set of field parameters (\ref{Q_parameters}) in $\omega_R$
units. We see strong equivalence for the steady-state distribution
of laser cooled $^{24}Mg$ ($\omega_R/\gamma \simeq 1100$), $^{40}Ca$
($\omega_R/\gamma \simeq 32.3$), and $^{88}Sr$ ($\omega_R/\gamma
\simeq 0.635$) atoms at the narrow-line $^1S_0\to^3P_1$ optical
transitions that confirm the scalability principle
(\ref{Q_parameters})- (\ref{en_q}) of laser cooling on narrow-line
optical transitions.

First of all, notice sharp peaks in the momentum distribution at
$p=\pm hk$. These peaks represent the effects of velocity-selective
coherent population trapping
\cite{Aspect88,Aspect1989,Prudnikov2003} for the $\Lambda$ scheme of
the families (\ref{w_functionP}) with $p_0 = \hbar k$
(Fig.\ref{P_levels}(a)) and was first demonstrated in a two-level
system for $2^3S_1 \to 3 ^3P_2$ optical transition in metastable
helium \cite{olshanii} with $\omega_R/\gamma \simeq 0.22$.

\begin{figure}[!t]
\centerline{\includegraphics[width=3.1 in]{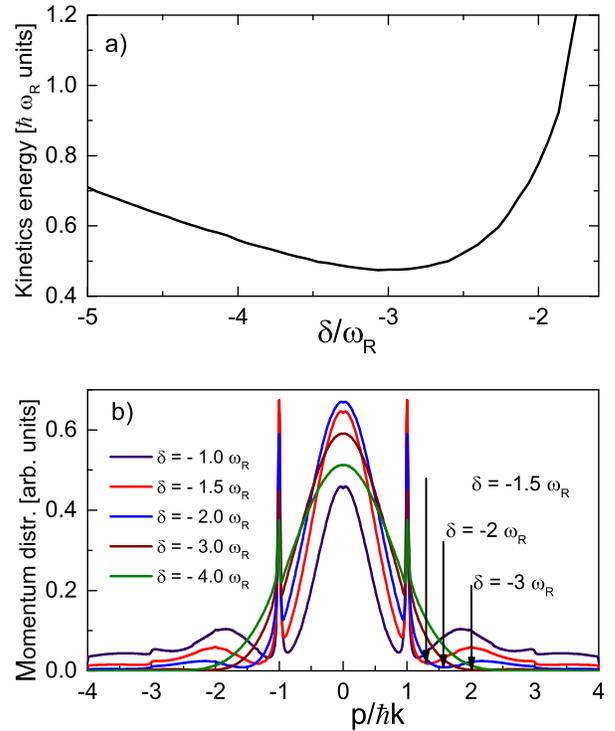}} \caption{\em
The kinetic energy (a) and the steady-state momentum distribution of
laser cooled atoms (b) for different detunings and small Rabi
$\Omega_0 = 0.64 \,\omega_R$.} \label{sub_resonances}
\end{figure}

The numerical simulation Fig.\ref{sub_resonances}(a) shows the
minimum kinetic energy of laser cooling on narrow lines
(\ref{quantum}) in low intensity standing wave is reached for
detuning about
\begin{equation}\label{delta_opt}
\delta^* \simeq -3 \,\omega_R\,
\end{equation}
that is far beyond optimal condition of laser cooling in
semiclassical limit and close to one was shown in \cite{castin1989}
for $\sigma_+-\sigma_-$ light field configuration. The momentum
distribution is significantly nongaussian function for all
considered range of detuning Fig.\ref{sub_resonances}(b).

The possible qualitative explanation of optimal value of detuning
(\ref{delta_opt}) can be given as follows. For a low field intensity
the population of the excited state is negligibly small.  In this
case, the atom distribution in the momentum space can be represented
as a set of families (\ref{w_functionP}) with nonzero amplitudes
$\alpha_n(p_0)$ near $p_0 = 0$ (i.e. $|g,p=0\rangle$ and $|g,p=\pm 2
\hbar k\rangle$ states), that results in a main peak in the momentum
distribution with $p=0$ and two side peaks at $p=\pm 2 \hbar k$
(Fig. \ref{sub_resonances}(b)). To obtain a minimum kinetic energy,
it is necessary to suppress the amplitudes of these side peaks. The
depopulation of $|g, p = \pm2\hbar k \rangle$ states is reached for
detuning $\delta_p$ providing resonance for transitions $|g, p =
\pm2\hbar k \rangle \to |e, p = \pm \hbar k \rangle$, i.e. for
$\delta_p = \delta + 3\omega_R = 0$ according to (\ref{deltaP}), see
Fig.\ref{P_levels}(b). Thus the detuning $\delta = -3\omega_R$
results effective suppression of side peaks of momentum distribution
at $p=\pm 2\hbar k$ Fig.\ref{P_levels}(b) and effective population
of $|g,p\rangle$ state around $p=0$.

Finally, an important question is the time evolution of atomic
distribution and cooling time. In the quantum regime being
considered (\ref{quantum}), the atom cooling rate can be represented
in equivalent form that is different form the semiclassical case
(\ref{semiclassic}). The scaling low for cooling rate on narrow
lines can be written for dimensionless time $t/t_Q$ in the following
form:
\begin{equation}
t_Q = \gamma^{-1} \cdot \tau_Q\left(\Omega/\omega_R,\,
\delta/\omega_R \right)\,
\end{equation}
where time $t_Q$ scales in inverse $\gamma$ units apart from the
semiclassical definition (\ref{semtime}) and a function of set of
dimensionless parameters $\Omega/\omega_R$ and $\delta/\omega_R$
that define scaling law in quantum regime of laser cooling
(\ref{mom_Q}).

\begin{figure}[!t]
\centerline{\includegraphics[width=3.8 in]{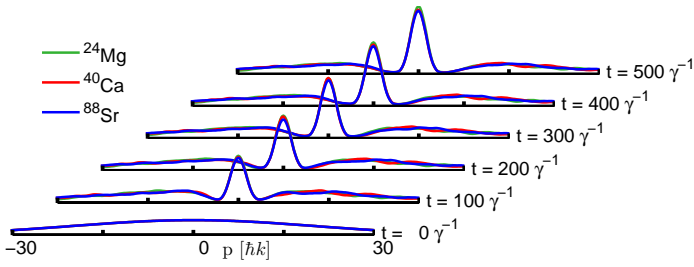}} \caption{\em
Time evolution of the momentum distribution of laser cooling various
atoms for equivalent time in $\gamma^{-1}$ units. Light field
parameters: $\Omega_0 = 2\omega_R$, detuning $\delta  = -3
\omega_R$, the initial momentum distribution is Gaussian function
with $p_{\sigma} = 20 \hbar k$.} \label{MC}
\end{figure}

Fig.\ref{MC} shows the time evolution of the momentum distribution
obtained by the quantum Monte-Carlo method \cite{Molmer1993} with
averaging over 3000 trajectories. Here we consider the laser cooling
from an initial momentum distribution $p_{\sigma} = 20 \hbar k$
corresponding the initial temperature of Mg atoms, $T \simeq 1.5
mK$, that is, the laser cooling temperature in a MOT with use of the
strong dipole optical transition $^1S_0 \to ^1P_1$ \cite{Rasel2012},
which can be treated as an initial stage of laser cooling for
further subrecoil cooling temperatures with the use of the
narrow-line optical transition.

\begin{figure}[!t]
\centerline{\includegraphics[width=2.5 in]{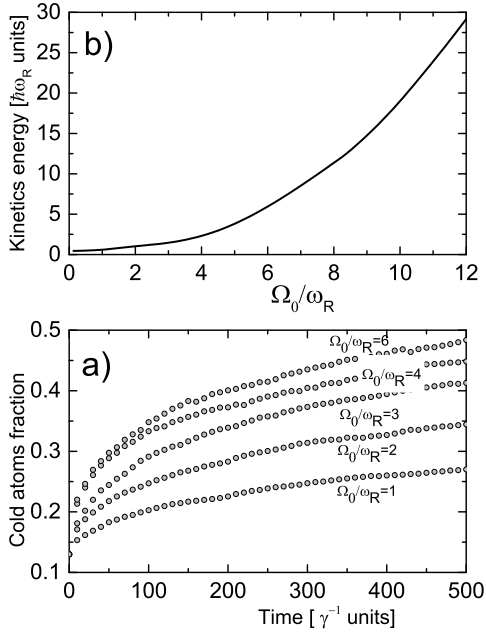}} \caption{\em
The steady-state kinetics energy of atoms (a) and as function of
Rabi in $\omega_R$ units. Detuning $\delta = -3 \omega_R$ and the
cold atom fraction with momentum $|p|<3 \hbar k$ time evolution in
$\gamma^{-1}$ units (b)} \label{temp_time}
\end{figure}

As was discussed above, the slow cooling rate with use of
narrow-line optical transition is the main obstacle for realization
of effective laser cooling in monochromatic light. The different
modification like cooling with broadband light and ``quenching
cooling'' techniques we considered
\cite{Ertmer1989,holberg2001,Rasel2001,Rasel2003}. However in
``scaling law'' for considered quantum the regimes of laser cooling
significantly different from the semiclassical one. In particular,
the field dependance in semiclassical regime $\Omega/\gamma$ changes
to $\Omega/\omega_R$ that gives possibility to use significantly
stronger laser field intensity without increasing much the
steady-state temperature of cold atoms Fig.\ref{temp_time}(a). Thus
the ``power broadening'' allows to significantly increase the
cooling rate without much lose in the temperature. The cooling time
can be considered by analyzing the evolution of the cold atom
fraction. The evolution of the cold atom fraction with the momentum
$|p|<3 \hbar k$ is shown in Fig.\ref{temp_time}(b). The cold atom
fraction grows fast for the times $t\sim 100 \gamma^{-1}$ and enough
large Rabi $\Omega_0 = 6 \omega_R$. The stronger light field results
in faster evolution but in a larger steady-state kinetic energy of
the atoms (Fig.\ref{temp_time}(b)).

In attempts to reach deep laser cooling of atoms with nondegenerate
ground states, such as $^{24}Mg$, $^{40}Ca$, and $^{88}Sr$ the
researchers tried to use the narrow-line optical transitions.
However, due to the limitation of the semiclassical description of
the standard Doppler theory of laser cooling, the optimal light
field parameters differ significantly. This may not lead to the
expected subrecoil laser cooling temperatures in experiments with
light field detuning and intensity extrapolated from standard the
semiclassical theory. The laser cooling at the narrow-line optical
transition requires \ special study we perform here. We have shown
the equivalence of laser cooling of various atoms with $\omega_R\geq
\gamma$ for a set of dimentionless light field parameters
($\Omega/\omega_R$ and $\delta/\omega_R$) scaled in recoil frequency
units that represent a scaling low in laser cooling on narrow-line
optical transition. In introdused dimetionless units the momentum
distribution of various atoms are equivalent for momentum scaled in
$\hbar k$ units. We have demonstrated that in standing wave the
kinetic energy of cold atoms reached it minimum for detuning close
to $\delta \simeq -3 \omega_R$, i.e. is far from the Doppler theory
of laser cooling.

The cooling time with the use of narrow-line optical transition
scales in the $\gamma^{-1}$ units differ from the semiclassical
limit (cooling time scales in the $\omega_R^{-1}$ units) and
efficient cooling rate can be achieved with ``power broadening'' for
the Rabi of few recoil frequency.

The research was supported by RSF (project No. 16-12-00054).


\end{document}